\documentclass[11pt,a4paper,oneside]{article}
\usepackage[T1]{fontenc}
\usepackage[ansinew]{inputenc}
\usepackage[english]{babel}
\usepackage{amsfonts}
\usepackage{amsmath}
\usepackage{bm}
\usepackage{array}
\usepackage{amsthm}
\usepackage{amssymb}
\usepackage{graphicx}
\usepackage{braket}
\usepackage{verbatim}
\usepackage[table]{xcolor}
\usepackage{caption}
\usepackage{cite}
\usepackage{textcomp}
\usepackage{url}
\usepackage{hyperref}
\usepackage{tikz}
\usepackage{pgfplots}
\usepackage{tensor}
\usepackage{mathtools}
\hypersetup{
    colorlinks=true,
    linkcolor=black,
    citecolor=black
    }
\usepackage{multicol}
\newcommand{\be}{\begin{equation}}
\newcommand{\ee}{\end{equation}}
\definecolor{pinegreen}{rgb}{0.0, 0.47, 0.44}

\theoremstyle{definition}

\theoremstyle{remark}

\newcommand\numberthis{\addtocounter{equation}{1}
\tag{\theequation}}
\title{\bf Quasinormal modes for coherent quantum black holes} %
\author{
Tommaso~Antonelli$^{1}$,
Andrea~Giusti$^{2}$, 
Roberto~Casadio$^{3,4}$,
Lavinia~Heisenberg$^{5}$
\\
\\
$^1${\em Department of Physics and Astronomy, University of Sussex}\\ 
{\em Brighton, BN1 9QH, United Kingdom}
\\
\\
$^2${\em Department of Physics \& Astronomy, Bishop's University}
\\
{\em 2600 College Street, Sherbrooke, Qu{\'e}bec, Canada J1M 1Z7} 
\\
\\
$^3${\em Dipartimento di Fisica e Astronomia, Universit\`a di Bologna}
\\
{\em via Irnerio~46, 40126 Bologna, Italy}
\\
\\
$^4${\em I.N.F.N., Sezione di Bologna, I.S.~FLAG}
\\
{\em viale B.~Pichat~6/2, 40127 Bologna, Italy}
\\
\\
$^5${\em Institute for Theoretical Physics, Heidelberg University}
\\
{\em Philosophenweg 16, 69120 Heidelberg, Germany} 
}
\begin{document}
\maketitle
\begin{abstract}
\noindent 
Coherent quantum black holes are quantum geometries obtained by means of a mean-field-like approach
to the gravitational interaction.
This procedure attenuates the classical spacetime singularities of general relativity by replacing them with
integrable singularities in the quantum-corrected geometry.
After discussing some relevant observables for a novel geometry for spherically symmetric black holes,
we investigate the quasinormal modes spectrum of scalar, electromagnetic, and gravitational fields
for the proposed model.
The results indicate potential deviations from general relativity, the magnitude of which is gauged by the value
of the ultraviolet regulator of the model (physically identifiable as a matter core).
Observations of the ringdown phase in black hole mergers could help detect such deviations.
\end{abstract}
\vspace{1cm}
{Preprint: ET-0142A-25}
\newpage
\section{Introduction}
\label{sec-1}
Coherent quantum black holes are emerging quantum geometries for which the
metric tensor is obtained, following a mean-field-like approach to the gravitational
interaction, which is expected to hold also in the strong-coupling regime. 
\par
Let us consider a static spherically symmetric line element
\begin{equation}
\label{eq:linelement}
ds^2
=
-f(r)\,dt^2 + \frac{dr^2}{f(r)}+r^2 \, d\Omega^2
\ ,
\qquad
f
=
1+2\,V(r)
\ ,
\end{equation}
where $d\Omega^2$ denotes the line element on the $2$-sphere, $r$ is the areal radius,
and we shall refer to $V=V(r)$ as potential function, although no weak-field approximation is
involved in the construction.~\footnote{For the sake of convenience, we set
Newton's gravitational constant and the reduced Planck constant to unity.}
\par
If we consider a classical reference geometry (from general relativity) given in terms of the
potential $V=V_{\rm c}(r)$, then one can think of this geometry as the expectation value of a
metric tensor operator $\widehat{g}_{ab}$ over a suitable quantum state $\ket{g}$,
that is
\begin{equation}
\label{eq:QuantumState}
\braket{g|\widehat{V}|g}
=
V_{\rm c} (r)
\ .
\end{equation}
The underlying idea of the coherent state approach~\cite{Casadio:2021eio} is to
describe $\widehat{V}$ as a free massless scalar field which is meant to represent
the non-perturbative collective behaviour of the only gravitational degree of freedom
required to reproduce the solution $V_{\rm c}$ from the quantitation of the Einstein
theory.
Hence, within this approach, a certain classical configuration $V=V_{\rm c}(r)$ can be
realised if there exists a coherent state $\ket{g}$ of the field $\widehat{V}$
such that Eq.~\eqref{eq:QuantumState} holds.
The total occupation number $N$ enters the normalisation of the coherent state $\ket{g}$
and measures the distance of $\ket{g}$ from the vacuum $\ket{0}$, the latter being
understood as the state devoid of any gravitational and matter excitation~\cite{Casadio:2021eio}.
\par
Taking the Schwarzschild metric with
\be
V_{\rm c}
=
-\frac{M}{r}
\ee
as the reference classical geometry and characterising the associated coherent state
in momentum space, one finds~\cite{Casadio:2021eio}
\begin{equation}
N
=
4\,M^2 \int_0^\infty \frac{dk}{k}
\ ,
\end{equation}
which diverges both in the ultraviolet (UV) and in the infrared (IR) regimes.
This means that the exact Schwarzschild black hole cannot be realised within
the coherent state approach, and a regularisation is required in order to find a coherent
state that closely (yet, not exactly) reproduces the Schwarzschild metric.
As discussed in Ref.~\cite{Feng:2024nvv}, this can be achieved by introducing an IR
cut-off accounting for the finite lifetime of a black hole, say $\tau \sim R_\infty \equiv k_{\rm IR}^{-1}$,
and a Gaussian regulator related to a quantum core of finite size $R_{\rm s}\equiv k_{\rm UV}^{-1}$.
This yields
\begin{equation}
N
=
4\,M^2
\int_{R_\infty^{-1}}^\infty
\frac{e^{-{(R_{\rm s}\,k)^2}/{2}}}{k}
\,dk
\simeq  
4\,M^2\,
\log\left( \frac{R_\infty}{R_{\rm s}}\right)
\ ,
\end{equation}
under the assumption that $R_{\rm s} \ll R_{\infty}$.
This regularisation, when implemented at the level of the potential, yields a
quantum-corrected Schwarzschild geometry defined by the quantum
potential~\cite{Feng:2024nvv}
\begin{equation}
\label{eq:QuantumPotential}
V_{\rm q}
=
-\frac{M}{r} \, {\rm erf}\left( \frac{r}{R_{\rm s}} \right)
\ ,
\end{equation}
where ${\rm erf} (z)$ denotes the error function.
Clearly, Eq.~\eqref{eq:QuantumPotential} tells us that the size of the quantum
core $R_{\rm s}$ gauges the strength of the corrections to the classical
Schwarzschild potential, which is (formally) obtained for $R_{\rm s} \to 0^+$.
\par
A similar analysis has been performed for electrically charged black
holes~\cite{Casadio:2022ndh}, where the regularisation of the UV divergence
was carried out by means of a sharp UV cut-off as in Ref.~\cite{Casadio:2021eio}
for the electrically neutral case.
The problem with this regularisation is that it introduces (spurious) osculations
in the quantum potential, due to the fact that the error function in
Eq.~\eqref{eq:QuantumPotential} is replaced by a sine-integral function.
This feature is removed by the smoother regularisation adopted in Ref.~\cite{Feng:2024nvv}.
Notably, the Gaussian regulator used in Ref.~\cite{Feng:2024nvv} 
does not affect the main interior features of the quantum core
originally derived in Ref.~\cite{Casadio:2021eio}.
Specifically, our quantum-corrected black hole contains an integrable singularity at its center,
as we shall discuss in Section~\ref{sec-2}.
\par
Quasinormal modes are the characteristic oscillations of perturbations propagating
on a black hole background geometry.
If we denote by $\psi$ the propagating physical degree of freedom associated to a
given perturbed field on a metric of the form~\eqref{eq:linelement}, one can express
its time dependence as
\begin{equation}
\psi
\sim
e^{-i\,\omega\, t}
\ ,
\end{equation}
with $\omega$ being the quasinormal mode frequency of the black hole.
The peculiarity of perturbation theory on a black hole background is that
the system is dissipative (due to the presence of an event horizon) and the
frequencies $\omega$ for modes that are purely outgoing at infinity are therefore
complex, 
\begin{equation}
  \omega
  =
  \omega_{\rm R}+i\,\omega_{\rm I}
  \ , 
  \qquad
  \omega_{\rm I}<0
  \ ,
\end{equation}
with the imaginary part yielding the timescale of decay for the mode.
(Excellent reviews of black hole quasinormal modes are, for instance,
Refs.~\cite{QNMberti, QNMkokkotas, WKBiyerwill}.)
\par
For the purposes of this work, we will focus solely on perturbations outside the event horizon,
since this is the most relevant region of spacetime for astrophysical observations,
and we will consider perturbations of three different types of fields:
scalar, electromagnetic, and gravitational.
The scalar field represents a simple toy model that contains the relevant features
for a stability analysis of the solution, without the mathematical complications of
gauge fields.
The electromagnetic and gravitational fields are of course more interesting
from a phenomenological perspective.
\par
This work is organised as follows:
in Section~\ref{sec-2} we review the quantum-corrected Schwarzschild black hole
obtained within the coherent state approach in Ref.~\cite{Feng:2024nvv};
in Section~\ref{sec-3} we estimate the quasinormal mode frequencies for
this geometry {\em via} the semi-analytical WKB procedure proposed in
Ref.~\cite{WKBiyerwill};
lastly, a discussion and outlook on future research is presented in Section~\ref{sec-4}.
\section{Quantum-corrected Schwarzschild geometry}
\label{sec-2}
Let us consider the line element in Eq.~\eqref{eq:linelement} with $f=f_{\rm q}=1 + 2\,V_{\rm q} (r)$
and the quantum potential $V_{\rm q}$ defined in Eq.~\eqref{eq:QuantumPotential}.
Recalling that the line element of a static spherically symmetric spacetime, with vanishing 
anomalous redshift, can be parametrised in terms of the Misner-Sharp-Hernandez mass $m=m(r)$
as (see e.g.~Ref.~\cite{Faraoni:2015ula})
\begin{equation}
f_{\rm q}
=
1 - \frac{2\,m(r)}{r}
\ ,
\end{equation}
one can easily infer that
\begin{equation}
m
=
-r\,V_{\rm q} (r)
=
M \, {\rm erf} \left( \frac{r}{R_{\rm s}} \right)
\ .
\end{equation}
Taking advantage of the discussion in Ref.~\cite{Casadio:2023iqt} for general static
spherically symmetric spacetimes, one can easily compute the Kretschmann scalar
${\cal K}=R^{\mu \nu \alpha \beta}\, R_{\mu \nu \alpha \beta}$ and the effective
energy-momentum tensor for our quantum-corrected Schwarzschild geometry.
Interestingly, since 
\begin{equation}
m
= 
\frac{2\,M\,r}{\sqrt{\pi}\,R_{\rm s}}
+
\mathcal{O} \left[\left( \frac{r}{R_{\rm s}} \right)^ 3\right]
\ ,
\end{equation}
for $r \to 0^+$, it is easy to see that
\begin{eqnarray}
{\cal K}
&\!\!=\!\!&
\mathcal{O}\left( \frac{1}{r^4}\right)
\ ,
\label{eq:K}
\\
\rho
&\!\!=\!\!&
-p_r
=
\frac{m'(r)}{4\,\pi\,r^2}
=
\mathcal{O}\left(\frac{1}{r^2}\right)
\ , 
\label{eq:rho}
\\
p_\theta
&\!\!=\!\!&
p_\phi
=
\frac{m''(r)}{8\,\pi\, r}
=
\mathcal{O}\left( 1\right)
\
\label{eq:p}
\ ,
\end{eqnarray}
for $r \to 0^+$, where $\rho$, $p_r$, $p_\theta$, and $p_\phi$ denote the effective energy
density and pressures associated to the quantum-corrected metric.
\par
From Eqs.~\eqref{eq:K}-\eqref{eq:p} one can easily conclude that the spacetime still
contains a curvature singularity at $r=0$, but the divergence of ${\cal K}$ is much
milder than Schwarzschild's ${\cal K}_{\rm Sch} =\mathcal{O}( 1/r^6)$.
Furthermore, although the effective energy-momentum tensor has divergences at $r=0$,
the volume integrals of the effective energy density and pressures are finite in any neighbourhood
of $r=0$, i.e.
\be
\left|4\,\pi \int_0 ^\epsilon \rho(\bar{r}) \, \bar{r}^2\, d\bar{r} \right|
<
\infty 
\qquad
\mbox{and}
\qquad
\left|4\,\pi \int_0^\epsilon p_i (\bar{r}) \, \bar{r}^2 \,d\bar{r} \right| < \infty
\ , 
\quad \forall \epsilon > 0
\ ,
\ee
where $i=(r,\theta,\phi)$.
In other words, we find that our quantum corrected geometry has an
{\em integrable singularity\/} at $r=0$ (see Ref.~\cite{Lukash:2013ts} for more details).
\par
The existence of an integrable singularity at the center of a coherent quantum black hole
was already pointed out in Refs.~\cite{Casadio:2021eio,Casadio:2022ndh},
although with a different regularisation procedure.
Here we have shown that replacing the sharp UV cut-off with a Gaussian regulator
does not affect the behaviour of the quantum corrected metric near $r=0$.
\par
For a spacetime metric~\eqref{eq:linelement}, the location of horizons is given by
the solutions of the equation
\begin{equation}
\label{eq:horizons}
    f_{\rm q} (r_{\rm H})= 1 + 2V_{\rm q} (r_{\rm H}) =0
    \ ,
\end{equation}
equivalent to $r_{\rm H} = 2 \,m(r_{\rm H})$.
Then, it is easy to see that Eq.~\eqref{eq:horizons} admits an event horizon if and only if 
\begin{equation}
    R_{\rm s}
    <
    \frac{4\,M}{\sqrt{\pi}}
    =
    \frac{2\, R_{\rm Sch}}{\sqrt{\pi}}
    \approx
    1.13\,R_{\rm Sch}
    \ ,
\end{equation}
where $R_{\rm Sch} = 2\,M$ is the Schwarzschild radius.
If we further require the quantum core to be hidden inside the actual event horizon at $r_{\rm H}$,
i.e.~$R_{\rm s} < r_{\rm H}$, we have the additional restriction on the size of the core
\begin{equation}
R_{\rm s}
<
{\rm erf}(1) \, R_{\rm Sch}
\approx
0.84 \, R_{\rm Sch}
\ .
\label{BHb}
\end{equation}
The above bound implies that a core of size $R_{\rm s}=R_{\rm Sch}$ could be considered as
a kind of black hole {\em mimicker\/} (like the {\em gravastars\/}~\cite{Mazur:2004fk}).
\par
As discussed in Ref.~\cite{Urmanov:2024qai}, several physical observable can be computed
in order to investigate deviations from general relativity in the context of coherent quantum
geometries.
For instance, a rather straightforward problem is to compute the location of the 
photon ring $R_\gamma$, defined as the solution to the equation~\cite{Urmanov:2024qai}
\begin{equation}
  R_\gamma\,f'_{\rm q}(R_\gamma)-2\,f_{\rm q}(R_\gamma)
  =
  0
  \  ,
\end{equation}
which ultimately represents the lowest bound for stable orbits.
\par
Another interesting observable is the critical impact parameter for massless particles
$b_{\rm c}$, which is defined as~\cite{Urmanov:2024qai}
\begin{equation}
b_{\rm c}
=
\frac{R_\gamma}{\sqrt{f_{\rm q}(R_\gamma)}}
\ ,
\end{equation}
and represents the minimal ratio of the angular momentum over the energy of the
massless particle below which the particle falls into the black hole. 
\par
\begin{figure}
  \centering
  \captionsetup{width=0.9\textwidth, font=footnotesize}
  \includegraphics[width=0.8\textwidth]{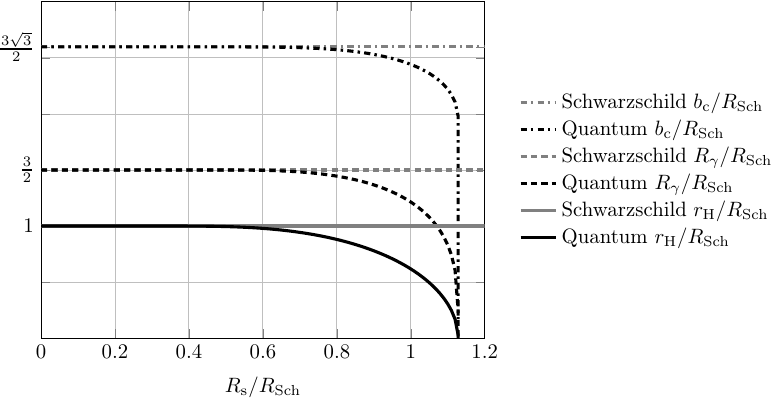}
  \caption{Values of $r_{\rm H}$, $R_\gamma$, and $b_{\rm c}$ in units of $R_{\rm Sch}$
  as a function of the size of the quantum core $R_{\rm s}$.}
  \label{fig:allinone}
\end{figure}

In Fig.~\ref{fig:allinone} we display the magnitude of the deviations of $r_{\rm H}$,
$R_\gamma$, and $b_{\rm c}$ from the corresponding values for the Schwarzschild
solution (for which $r_{\rm H}=R_{\rm Sch}$, $R_\gamma=3\,R_{\rm Sch}/2$,
and $b_{\rm c}=3\sqrt{3}R_{\rm Sch}/2$) as we vary the size of the quantum core
$R_{\rm s}$.
\section{Quasinormal modes}
\label{sec-3}
Quasinormal modes of a static spherical symmetric black hole, for scalar perturbations,
can be computed by considering a scalar field $\Phi$ satisfying the Klein-Gordon equation
\begin{equation}
\label{eq:KG}
\left(
\Box
-m^2
\right)
\Phi
=
0
\ .
\end{equation}
It is convenient to decompose the scalar field in Fourier modes and spherical harmonics $Y_\ell^m=Y_\ell^m(\theta,\varphi)$,
\begin{equation}
\Phi
=
\int d\omega\, 
\sum_{\ell m}\,
e^{-i\,\omega\, t}\,
\Phi_{\omega \ell m}(r,\theta,\varphi)
\ .
\end{equation}
where~\footnote{These modes decouple in the static and spherically symmetric
background~\eqref{eq:linelement}.}
\begin{equation}
\Phi_{\omega \ell m}
=
F_{\omega \ell m}(r) \,Y_{\ell}^m(\theta,\varphi)
\ ,
\end{equation}
with $\ell=0,1,\ldots$ and $m=-\ell,\ldots,\ell$ as usual.
Plugging this ansatz into Eq.~\eqref{eq:KG}, and adopting the field redefinition
\begin{equation}
\psi_0(r)
\equiv
r\,F_{\omega \ell m}(r)
\ ,
\end{equation}
turns Eq.~\eqref{eq:KG} into 
\begin{equation}
\label{eq:schrodinger_scalar}
\frac{d^2\psi_0}{dr_*^2}
+
\left[\omega^2-V_0(r)\right]
\psi_0
=
0
\ ,
\end{equation}
where $r_*$ denotes the tortoise coordinate, defined by the condition $dr_*=dr/f(r)$, and 
\begin{equation}
\label{eq:scalarpot}
V_0
=
f(r)
\left[\frac{\ell (\ell+1)}{r^2}+\frac{f'(r)}{r}+m^2\right]
\ .
\end{equation}
More general perturbations $\psi_j$ can be shown to satisfy similar Schr\"odinger-like equations
\begin{equation}
\label{eq:schrodinger}
\frac{d^2\psi_j}{dr_*^2}
+
\left[\omega^2-V_j(r)\right]
\psi_j
=0
\  ,
\end{equation}
where the potential $V_i=V_j(r)$ is uniquely defined for each propagating degree of freedom $\psi_j$.
\par
Quasinormal modes are solutions of Eq.~\eqref{eq:schrodinger} in the region outside the event horizon
with appropriate boundary conditions for their asymptotic expansions.
Assuming that the potential $V_j$ is finite for $r\to r_{\rm H}^+$ and $r\to\infty$, the function $\psi_j$
behaves as
\begin{equation}
\label{eq:asymptotics}
\psi_j(r_*)\sim
\begin{dcases}
Z_{\rm H}^{\rm (out)}\,e^{-i \,k_{\rm H}\, r_*}
+ 
Z_{\rm H}^{\rm (in)}\,e^{+i\, k_{\rm H}\, r_*}
\qquad
\text{for}
\quad
r\to r_{\rm H}^+
\\
\\
Z_{\infty}^{\rm (out)}\,e^{+i \,k_\infty \,r_*}
+
Z_{\infty}^{\rm (in)}\,e^{-i\, k_\infty\, r_*}
\qquad
\text{for}
\quad
r\to \infty
\ ,
\end{dcases}
\end{equation}
where
\be
k_{\rm H}^2
=
\omega^2
-
\lim_{r\to r_{\rm H}^+}V_j(r)
\ , 
\qquad
\text{Re}(k_{\rm H})>0
\ee
and
\be
k_\infty^2
=
\omega^2
-
\lim_{r\to \infty}V_j(r)
\ , 
\qquad
\text{Re}(k_\infty)>0
\ .
\ee
The coefficients $Z_{\rm H}^{\rm (out/in)}$ represent the amplitudes of outgoing and ingoing modes at the horizon,
respectively.
Similarly, $Z_{\infty}^{\rm (out/in)}$ are the amplitudes of modes escaping towards or falling from infinity.
Since no information can exit the horizon~\footnote{This strictly classical property of the horizon is further
analysed in the quantum theory in Ref.~\cite{Feng:2024nvv}.} and we are only interested in perturbations generated 
near the black hole, we assume
\begin{equation}
\label{eq:inmodes}
Z_{\rm H}^{\rm (in)}=Z_{\infty}^{\rm (in)}
=
0
\ .
\end{equation}
This completely characterizes the quasinormal mode frequencies $\omega$, which are found to be
quantized.
\par
\begin{figure}
  \centering
  \captionsetup{width=0.9\textwidth, font=footnotesize}
  \includegraphics[width=0.43\textwidth,height=0.3\textwidth]{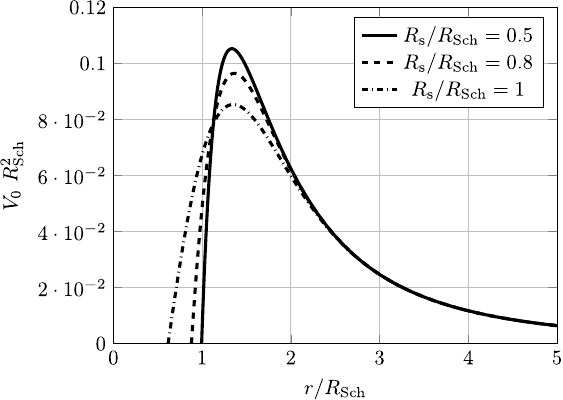}
  $\qquad$
   \includegraphics[width=0.43\textwidth,height=0.3\textwidth]{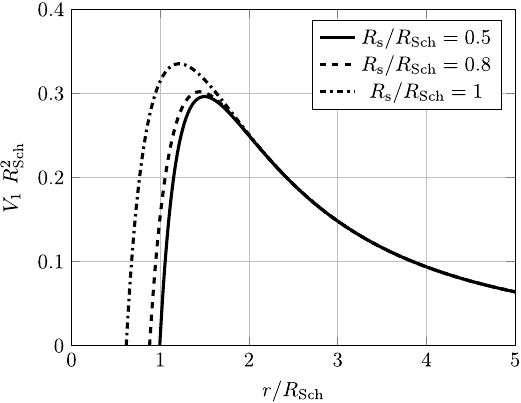}
   \\
   $\ $
   \\
    \includegraphics[width=0.43\textwidth,height=0.3\textwidth]{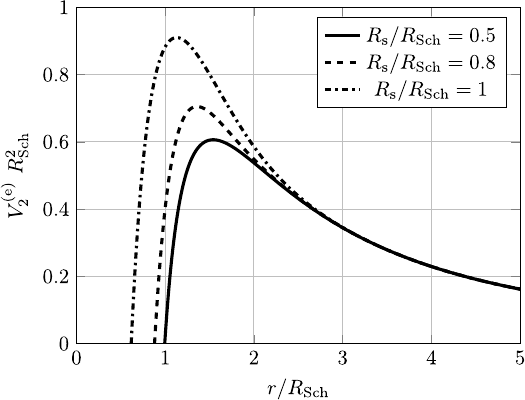}
    $\qquad$
     \includegraphics[width=0.43\textwidth,height=0.3\textwidth]{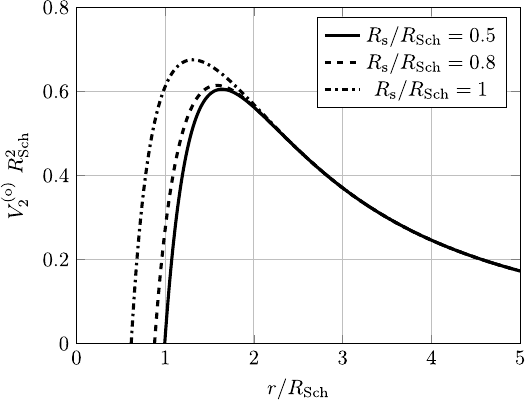}
  \caption{Potentials for the quasinormal modes for different values of $R_{\rm s}$: 
  $V_0$ for scalar perturbations (top left);
  $V_1$ for vector perturbations (top right);
  $V_2^{({\rm e})}$ for even tensor perturbations (bottom left);
  $V_2^{({\rm o})}$ for odd tensor perturbations (bottom left).}
  \label{fig:Vs}
\end{figure}

For a scalar field (spin $j=0$) with mass $m$, minimally coupled to gravity, the potential $V_0$ is given
in Eq.~\eqref{eq:scalarpot}.
For the electromagnetic field (spin $j=1$), minimally coupled to gravity, we have two propagating
degrees of freedom (even and odd), which share the same potential~\cite{EMcrispino}  
\begin{equation}
\label{eq:EMpot}
V_1(r)
=
f(r)\,
\frac{\ell (\ell+1)}{r^2}
\ .
\end{equation}
Lastly, the gravitational perturbations (spin $j=2$) have two propagating degrees
of freedom governed by different potentials, namely 
\begin{equation}
  \label{eq:oddgravpertpot}
  V^{(\text{o})}_2(r)
  =
  f(r)
  \left[\frac{\ell(\ell+1)}{r^2}-\frac{2\,(1-f(r))}{r^2}-\frac{f'(r)}{r}\right]
  \ ,
\end{equation}
for odd perturbations, and
\begin{equation}
\begin{split}
  \label{eq:evengravpertpot}
  V^{(\text{e})}_2(r)
  =
  &\,
  \frac{f(r)}{r^2\left[\ell(\ell+1)-2f(r)+rf'(r)\right]^2}\biggl\{2f(r)^2\Bigl[2\ell(\ell+1)+r^3f'''(r)\Bigr]\\ 
  &
  +\Bigl[r^2f'(r)^2-r^3f'(r)f''(r) +\ell(\ell+1)\,(\ell(\ell+1)+r^2f''(r))\Bigr]\Bigl[\ell(\ell+1)+rf'(r)\Bigr]\\
  &
  \hspace{0cm} -f(r)\Bigl[2\ell(\ell+1)r^2f''(r)-2r^4f''(r)^2+\ell(\ell+1)\left(4\ell(\ell+1)+r^3f'''(r)\right)\\
  &
  \hspace{1.3cm}+rf'(r)\left(2\ell(\ell+1)+2r^2f''(r)+r^3f'''(r)\right)\Bigr]\biggr\}
  \ 
\end{split}   
\end{equation}
for even perturbations (see Appendix~\ref{App-A} for more details).
The above potentials are displayed in Fig.~\ref{fig:Vs} for different values of the core radius.
One can easily see that the value at the peak of the potential for scalar perturbations decreases
for larger cores, whereas it increases for all the other (physically relevant) perturbations,
particularly for even gravitational perturbations.
The width of the potentials instead increases with the size of the core in all cases.
Note that the case $R_{\rm s}=R_{\rm Sch}$ exceeds the bound~\eqref{BHb} for 
the existence of the event horizon and corresponds to a black hole mimicker.
\par
We can compute the quasinormal mode frequencies in the WKB approximation
following the procedure proposed in Ref.~\cite{WKBiyerwill}.
In a nutshell, one begins by relating the asymptotic expansions at the horizon to the asymptotic
expansions at infinity via a Bogoliubov transformation
\begin{equation}
\begin{pmatrix}
Z_{\rm H}^{\rm (out)}
\\
Z_{\rm H}^{\rm (in)}
\end{pmatrix}
=
\begin{pmatrix}
M_{11} & M_{12}
\\
M_{21} & M_{22}
\end{pmatrix}
\begin{pmatrix}
Z_{\infty}^{\rm (out)}
\\
Z_{\infty}^{\rm (in)}
\end{pmatrix}
\ .
 \end{equation}
The element $M_{21}$ turns out to be proportional to $1/\Gamma(-n)$, with $n$ a non-negative
integer known as the overtone number expressed by the WKB expansion
\begin{equation}
\label{eq:WKBexp}
n+\frac{1}{2}
=
i\,\frac{\left(\omega^2-\bar{V}_j\right)}{\sqrt{-2\,\bar{V}_j''}}
-
\Lambda_2
-
\Lambda_3
-
\Lambda_4
-
\ldots
\ ,
\end{equation}
where $\Lambda_k$ is a function of $n$ and derivatives of $V_j$, taken with respect to $r_*$
and evaluated at $r=\bar{r}$ where the potential is maximum, $\bar{V}_j\equiv V_j(\bar{r})$.
The boundary conditions~\eqref{eq:inmodes} imply that $M_{21} = 0$, and Eq.~\eqref{eq:WKBexp} then
determines the spectrum $\omega=\omega_{n\ell}$ of the quasinormal mode frequencies by relating $\omega$ to the
overtone number $n$ and the parameters in the potential.
The number of terms in each $\Lambda_k$ grows very fast in size with the expansion,
so we will not report here the explicit form of these terms.
Further details can be found in Ref.~\cite{WKBiyerwill} for the order up to $\Lambda_3$,
in Ref.~\cite{WKBkonoplya} for the order up to $\Lambda_6$, and in Ref.~\cite{WKBmatopala}
for the order up to $\Lambda_{13}$.
\par
The WKB expansion, while providing a very straightforward method for calculating the quasinormal
modes frequencies, has some drawbacks.
First, there is no proof that the series in Eq.~\eqref{eq:WKBexp} converges, but including higher and
higher orders results in worse estimates of $\omega$ compared to purely numerical evaluations.
Moreover, there is no simple way of estimating the error for $\omega$ and determine the optimal
number of terms to include.
Lastly, the method yields sensible results only for small values of the overtone number $n$,
typically for $n \leq \ell$. 
\par
Regarding the first issue, we will employ Pad\'e approximants~\cite{WKBmatopala, WKBkonoplya2},
which generally improve the result for $\omega$ by replacing the polynomial approximant in
Eq.~\eqref{eq:WKBexp} with a rational function.
In particular, we will use the Pad\'e approximant of order  $[6/7]$.
Concerning the error estimate, since we are mostly interested in the order of magnitude of the
correction to the standard Schwarzschild values, we will simply not be concerned with it here
and restrict our calculation to cases with $n\leq \ell$. 
\par
The Mathematica code that implements the calculation of $\omega$, based on the library
by~\textsc{R.~A.~Konoplya et al.} presented in Ref.~\cite{WKBkonoplya2}, as well as the code
for the numerical computations in Fig.~\ref{fig:allinone}, is available upon request from
\href{mailto:t.antonelli@sussex.ac.uk}{t.antonelli@sussex.ac.uk}.
The results of the computations obtained with the WKB method are summarized in
Tables~\ref{T1}-\ref{T4}, where a question mark $(?)$ indicates 
cases for which the WKB method does not properly converge.
We will comment about these quasinormal frequencies in the next Section.
%
%
%
\begin{table}[h]
\begin{center}
  \def\arraystretch{1.4}
  \begin{tabular}{c||c|c|c|}
    & Schwarzschild & Quantum ($R_{\rm s}=0.5\,R_{\rm Sch}$) & Quantum ($R_{\rm s}=0.7\,R_{\rm Sch}$) \\ \hline \hline
    $n=0,\ \ell=0$ & \hspace{0.5cm} $0.221 - 0.210 \,i\  (?)$ & \hspace{0.7cm}$0.222 - 0.207\, i\  (?)$ & \hspace{0.7cm}$0.218 - 0.184\, i\  (?)$\\ \hline
    $n=0,\ \ell=1$ & $0.586 - 0.195 \,i$& $0.585 - 0.194\, i$ & $0.587 - 0.184\, i$\\ \hline
    $n=1,\ \ell=1$ & $0.529 - 0.612     \,i$& \hspace{0.7cm}$0.527 - 0.612\, i\  (?)$ & \hspace{0.7cm}$0.530 - 0.565\, i\  (?)$\\ \hline
    $n=0,\ \ell=2$ & $0.967 - 0.194 \,i$& $0.967 - 0.193\, i$ & $0.969 - 0.183\, i$\\ \hline
    $n=1,\ \ell=2$ & $0.928 - 0.591 \,i$& $0.927 - 0.589\, i$ & $0.928 - 0.554\, i$\\ \hline
    $n=2,\ \ell=2$ & $0.861 - 1.017 \,i$& \hspace{0.7cm}$0.870 - 1.013\, i\  (?)$ & \hspace{0.7cm}$0.865 - 0.948\, i\  (?)$\\ \hline
    $n=0,\ \ell=3$ & $ 1.351 - 0.193\,i$& $1.350 - 0.193\, i$ & $1.354 - 0.183\, i$\\ \hline
    $n=1,\ \ell=3$ & $ 1.321 - 0.585\,i$& $1.320 - 0.583\, i$ & $1.323 - 0.552\, i$\\ \hline
    $n=2,\ \ell=3$ & $ 1.267 - 0.992\,i$& $1.264 - 0.987\, i$ & $1.267 - 0.928\, i$\\ \hline
    $n=3,\ \ell=3$ & $ 1.198 - 1.422\,i$& \hspace{0.7cm}$1.199 - 1.401\, i\  (?)$ & \hspace{0.7cm}$1.212 - 1.316\, i\  (?)$\\ \hline
  \end{tabular}
  \caption{Scalar field (assuming, for simplicity, $m=0$).
  \label{T1}}
   \end{center}
 \end{table}
%
%
%
\begin{table}[h]
\begin{center}
  \def\arraystretch{1.4}
  \begin{tabular}{c||c|c|c|}
    & Schwarzschild & Quantum ($R_{\rm s}=0.5\,R_{\rm Sch}$) & Quantum ($R_{\rm s}=0.7\,R_{\rm Sch}$) \\ \hline \hline
    $n=0,\ \ell=1$ & \hspace{0.7cm}$0.497 - 0.185 \,i$\hspace{0.7cm} & $0.497 - 0.184\, i$ & $0.502 - 0.173\, i$\\ \hline
    $n=1,\ \ell=1$ & $0.429 - 0.587 \,i$& \hspace{0.7cm}$0.432 - 0.588\, i\  (?)$\ & \hspace{0.7cm}$0.446 - 0.542\, i\  (?)$\\ \hline
    $n=0,\ \ell=2$ & $0.915 - 0.190 \,i$& $0.915 - 0.189\, i$ & $0.919 - 0.179\, i$\\ \hline
    $n=1,\ \ell=2$ & $0.873 - 0.581 \,i$& $0.872 - 0.578\, i$ & $0.878 - 0.543\, i$\\ \hline
    $n=2,\ \ell=2$ & $0.802 - 1.003 \,i$& \hspace{0.7cm}$0.802 - 0.993\, i\  (?)$\ & \hspace{0.7cm}$0.817 - 0.932\, i\  (?)$\\ \hline
    $n=0,\ \ell=3$ & $1.314 - 0.191\,i$& $1.313 - 0.191\, i$ & $1.318 - 0.181\, i$\\ \hline
    $n=1,\ \ell=3$ & $1.283 - 0.579\,i$& $1.282 - 0.577\, i$ & $1.287 - 0.545\, i$\\ \hline
    $n=2,\ \ell=3$ & $1.228 - 0.984\,i$& $1.224 - 0.978\, i$ & \hspace{0.7cm}$1.231 - 0.919\, i\  (?)$\\ \hline
    $n=3,\ \ell=3$ & $1.156 - 1.413\,i$& \hspace{0.7cm}$1.152 - 1.393\, i\  (?)$\ & \hspace{0.7cm}$1.176 - 1.298\, i\  (?)$\\ \hline
  \end{tabular}
\caption{Electromagnetic field.
\label{T2}}
\end{center}
\end{table}
%
%
\begin{table}[h]
\begin{center}
  \def\arraystretch{1.4}
  \begin{tabular}{c||c|c|c|}
    & Schwarzschild & Quantum ($R_{\rm s}=0.5\,R_{\rm Sch}$) & Quantum ($R_{\rm s}=0.7\,R_{\rm Sch}$)\\ \hline \hline
    $n=0,\ \ell=2$ & $0.747 - 0.178 \,i$& $0.747 - 0.177 \,i$ & $0.750 - 0.167 \,i$ \\ \hline
    $n=1,\ \ell=2$ & $0.693 - 0.548 \,i$& $0.692 - 0.545 \,i$ & \hspace{0.7cm}$0.708 - 0.510 \,i\  (?)$ \\ \hline
    $n=2,\ \ell=2$ & \hspace{0.5cm} $0.604 - 0.953 \,i\  (?)$ & \hspace{0.7cm}$0.612 - 0.981 \,i\  (?)$ & \hspace{0.7cm}$0.636 - 0.897 \,i\  (?)$ \\ \hline
    $n=0,\ \ell=3$ & $1.199 - 0.185 \,i$& $1.199 - 0.185 \,i$ & $1.202 - 0.175\, i$ \\ \hline
    $n=1,\ \ell=3$ & $1.165 - 0.563\,i$& $1.164 - 0.560 \,i$ & $1.169 - 0.529\, i$  \\ \hline
    $n=2,\ \ell=3$ & $1.103 - 0.958\,i$& $1.100 - 0.952 \,i$ & \hspace{0.7cm}$1.109 - 0.895\, i\  (?)$  \\ \hline
    $n=3,\ \ell=3$ & $1.024 - 1.381\,i$& \hspace{0.7cm}$1.030 - 1.358 \,i\  (?)$ & \hspace{0.7cm}$1.057 - 1.292\, i\  (?)$  \\ \hline
  \end{tabular}
\caption{Gravitational field (odd perturbations).
\label{T3}}
\end{center}
\end{table}
%
%
%
\begin{table}[h]
\begin{center}
  \def\arraystretch{1.4}
  \begin{tabular}{c||c|c|c|}
    & Schwarzschild & Quantum ($R_{\rm s}=0.5\,R_{\rm Sch}$) & Quantum ($R_{\rm s}=0.7\,R_{\rm Sch}$)\\ \hline \hline
    $n=0,\ \ell=2$ & $0.747 - 0.178 \,i$ & $0.750 - 0.175 \,i$ & \hspace{0.7cm}$0.761 - 0.152\, i\  (?)$ \\ \hline
    $n=1,\ \ell=2$ & $0.693 - 0.548 \,i$ & \hspace{0.7cm}$0.689 - 0.541 \,i\  (?)$ & \hspace{0.7cm}$0.698 - 0.456 \,i\  (?)$\\ \hline
    $n=2,\ \ell=2$ & \hspace{0.5cm} $0.602 - 0.956 \,i\  (?)$ & \hspace{0.7cm}$0.571 - 0.957 \,i\  (?)$ & \hspace{0.7cm}$0.539 - 0.720 \,i\  (?)$ \\ \hline
    $n=0,\ \ell=3$ & $1.199 - 0.185 \,i$ & $1.200 - 0.183\, i$ & $1.218- 0.172\, i$ \\ \hline
    $n=1,\ \ell=3$ & $1.165 - 0.563\,i$ & $1.167 - 0.557\, i$ & \hspace{0.7cm}$1.192- 0.521\,i\  (?)$ \\ \hline
    $n=2,\ \ell=3$ & $1.103 - 0.958\,i$ & \hspace{0.7cm}$1.106 - 0.952\, i\  (?)$ & \hspace{0.7cm}$1.149 - 0.895\, i\  (?)$ \\ \hline
    $n=3,\ \ell=3$ & $1.024 - 1.381\,i$ & \hspace{0.7cm}$1.037 - 1.401\, i\  (?)$ &\hspace{0.7cm}$1.085 - 1.338\, i\  (?)$ \\ \hline
  \end{tabular}
\caption{Gravitational field (even perturbations).
\label{T4}}
\end{center}
\end{table}
\section{Discussion and outlook}
\label{sec-4}
In this work, we explored the phenomenology of the quantum-corrected Schwarzschild black hole
geometry derived in Ref.~\cite{Feng:2024nvv} from the coherent state quantisation of gravity.
In particular, we considered constraints on the size $R_{\rm s}$ of the core
for which this geometry represents a black hole, and computed numerically
the horizon radius, photon ring, and critical impact parameter.
We then compared these observables with those of the classical Schwarzschild
geometry.
\par
Following this preliminary investigation of the properties of the proposed geometry,
we derived the effective potentials for the Schr\"odinger-like equations governing scalar,
electromagnetic, and gravitational perturbations on this background.
In particular, the quasinormal mode frequencies $\omega=\omega_{n\ell}$ for these fields were computed
using the standard WKB approximation with appropriate boundary conditions. 
\par
We observe that the quasinormal frequencies in Tables~\ref{T1}-\ref{T4}
deviate only slightly (order of $0.5\%$) from the Schwarzschild values, for a core of size
$R_{\rm s} = 0.5 \, R_{\rm Sch}$.
However, the discrepancies become more pronounced for larger values of $R_{\rm s}$.
In particular, the imaginary part of $\omega$ is always smaller than the corresponding Schwarzschild expectation
in the computed ranges, and becomes of the order of $5\%$ smaller for $R_{\rm s} = 0.7 \, R_{\rm Sch}$.
This result agrees with the behaviour of the width of the potential shown in Fig.~\ref{fig:Vs} 
and indicates that the decay time of such modes become (at least up to about $5\%$) longer for
larger cores, which would affect the duration of ringdown signals during a merger
(see, e.g.~Ref.~\cite{Nobili:2025ydt}).
The real part of $\omega$, on the contrary, remains relatively unaffected, which means that
quasinormal oscillations do not depend significantly on the size of the inner core.
Such features could be investigated with the next generation of gravitational wave 
detectors~\cite{Abac:2025saz}.
\par
We expect even larger deviations for larger values of $R_{\rm s}$.
However, the WKB method begins to show its limits in this regime. 
At $R_{\rm s} = 0.7 \, R_{\rm Sch}$, the WKB expansion fails to converge reliably
for many modes, especially in the gravitational sector.
Moreover, we recall that the horizon exists only for inner cores of size
$R_{\rm s} \lesssim 0.84 \, R_{\rm Sch}$.
Larger cores correspond to black hole mimickers without an event horizon, for which
the existence of quasinormal oscillations will strongly depend on the boundary condition 
at the surface $r=R_{\rm s}$ (which we have not analysed here).
Thus, our analysis of the quasinormal frequencies can serve as a {\em preliminary estimate}
of the order of magnitude of deviations of this quantum-corrected geometry from the classical case.
More refined results would require fully numerical calculations, which 
is however beyond the scope of this work.
\subsection*{Acknowledgements}
The work of T.A.~is supported by a doctoral studentship of the Science and Technology
Facilities Council (training grant No.~ST/Y509620/1, project ref.~2917813).
A.G.~and R.C.~carried out this work in the framework of the activities of the
Italian National Group of Mathematical Physics [Gruppo Nazionale per
la Fisica Matematica (GNFM), Istituto Nazionale di Alta Matematica (INdAM)].
R.C.~is partially supported by the INFN grant FLAG.
\appendix
\section{Quasinormal modes potentials for the gravitational sector}
\label{App-A}
Scalar, electromagnetic and gravitational perturbations on the background
metric~\eqref{eq:linelement} decouple, since the coupling between different
perturbations is proportional to the stress-energy tensor of the background solution,
which is assumed to be generated by some other unspecified field.
\par
Compared to the scalar and electromagnetic fields, the gravitational perturbations
are more involved, the main technical difficulty being the fact that the background
has a non-vanishing stress-energy tensor which interacts non-trivially with the
gravitational perturbations.
We start from the decomposition of the metric into background, given in Eq.~\eqref{eq:linelement},
and perturbations,
\begin{equation}
 g_{\mu\nu}
 =
 \bar{g}_{\mu\nu}+h_{\mu\nu}
 \ ,
 \end{equation}
which yields the perturbations of the Einstein equations
\begin{equation}
\left.\frac{\delta}{\delta g_{\rho\sigma}}
\left(G^{\mu\nu}-8\,\pi\,T^{\mu\nu}\right)\right|_{\bar{g}}
h_{\rho \sigma}
=
0
\ .
\end{equation}
\par
In the $(t,r,\theta,\varphi)$ coordinates of the background, the gravitational perturbation
under spatial rotations $SO(3)$ transforms as
\begin{equation}
h_{\mu\nu}=
\begin{pmatrix}\\[-2ex]
      \tikz{\node[draw] {S}} & \tikz{\node[draw] {S}} & \tikz{\node[draw, minimum width=1.5cm] {V}} \\[1ex]
      \tikz{\node[draw] {S}} & \tikz{\node[draw] {S}} & \tikz{\node[draw, minimum width=1.5cm] {V}} \\[1ex]
      \tikz{\node[draw, minimum height=1.5cm] {V}} & \tikz{\node[draw, minimum height=1.5cm] {V}} &
      \tikz{\node[draw, minimum width=1.5cm, minimum height=1.5cm] {T}}
    \end{pmatrix}
\end{equation}
where S, V and T denote a scalar, vector and tensor behaviour, respectively. 
We thus can introduce the vector spherical harmonics
\begin{align}
  &Y^{(1)}_{\ell m, \,a}(\theta,\varphi)=\left(\partial_\theta Y_{\ell m}\,,\, \partial_\varphi Y_{\ell m}\right)\\
  &Y^{(2)}_{\ell m, \,a}(\theta,\varphi)=\left(\frac{1}{\sin\theta}\,\partial_\varphi Y_{\ell m}\,,\, -\sin\theta\,\partial_\theta Y_{\ell m}\right)
\end{align}
and the tensor spherical harmonics
\begin{align*}
  &Y^{(1)}_{\ell m,\,ab}(\theta,\varphi)=
  \begin{pmatrix}
    \partial_\theta^2\, Y_{\ell m} & (\partial_\theta-\cot\theta)\partial_\varphi Y_{\ell m}\\[1.5ex]
    * & (\partial_\varphi^2+\sin\theta\cos\theta\,\partial_\theta)Y_{\ell m}
  \end{pmatrix}\numberthis\\
  &Y^{(2)}_{\ell m,\,ab}(\theta,\varphi)=
  \begin{pmatrix}
    Y_{\ell m} & 0\\[1ex]
    * & \sin^2\theta \ Y_{\ell m}
  \end{pmatrix}\numberthis\\
  &Y^{(3)}_{\ell m,\,ab}(\theta,\varphi)=
  \begin{pmatrix}
    \begin{aligned}
      \frac{1}{\sin\theta}(\partial_\theta-\cot\theta)\partial_\varphi Y_{\ell m}
    \end{aligned} & \begin{aligned}
      \frac{1}{2}\left(\frac{1}{\sin\theta}\partial_\varphi^2+\cos\theta\,\partial_\theta-\sin\theta\,\partial_\theta^2\right)\hspace{-1mm}Y_{\ell m}
    \end{aligned}\\[3ex]
    * & -\sin\theta(\partial_\theta-\cot\theta)\partial_\varphi Y_{\ell m}
  \end{pmatrix}
  \ .
  \numberthis
\end{align*}
The perturbation $h_{\mu\nu}$ correspondingly decomposes into even and odd parts under parity,
\begin{align*}
  \label{eq:decompositiongraveven}  
  &h^{(\text{e})}_{\mu\nu}(t,r,\theta,\varphi)
  =
  e^{-i\,\omega\, t}
  \begin{pmatrix}
    f(r) H_0(r) Y_{\ell m} & i\omega\,H_1(r) Y_{\ell m} & i\omega\,F_0(r) \,Y^{(1)}_{\ell m,\,a}\\
    * & 
    \begin{aligned}
      \frac{1}{f(r)}H_2(r) Y_{\ell m}
    \end{aligned} 
    & F_1(r) \,Y^{(1)}_{\ell m,\,a}\\
    * & * & r^2\left(G(r)Y^{(1)}_{\ell m,\,ab}+K(r)Y^{(2)}_{\ell m,\,ab}\right)
  \end{pmatrix}\numberthis\\
  \label{eq:decompositiongravodd}
  &h^{(\text{o})}_{\mu\nu}(t,r,\theta,\varphi)
  =
  e^{-i\,\omega\, t}
  \begin{pmatrix}
    0 & 0 & i\omega \,h_0(r) \,Y^{(2)}_{\ell m,\,a}\\
    * & 0 & h_1(r) \,Y^{(2)}_{\ell m,\,a}\\
    * & * & r^2\, h_2(r)\, Y^{(3)}_{\ell m,\,ab}
  \end{pmatrix}
  \ .
  \numberthis  
\end{align*}
Now, the perturbation of the Einstein tensor is readily obtained:    
\begin{align*}
 \left.\frac{\delta G_{\mu\nu}}{\delta g_{\rho\sigma}}\right|_{\bar{g}} h_{\rho \sigma}=\frac{1}{2}\Bigl(&\bar{\nabla}_\rho \bar{\nabla}_\mu \tensor{h}{_\nu^\rho}+\bar{\nabla}_\rho \bar{\nabla}_\nu \tensor{h}{_\mu^\rho}-\bar{\nabla}_\mu \bar{\nabla}_\nu \tensor{h}{_\rho^\rho}-\bar{\nabla}_\rho\bar{\nabla}^\rho h_{\mu\nu}\\
  &-\bar{R}h_{\mu\nu}+\bar{g}_{\mu\nu}\bar{\nabla}_\rho\bar{\nabla}^\rho \tensor{h}{_\sigma^\sigma}-\bar{g}_{\mu\nu} \bar{\nabla}_\rho \bar{\nabla}_\sigma h^{\rho\sigma}+\bar{g}_{\mu\nu} \bar{R}_{\rho\sigma}h^{\rho\sigma} \Bigr)\numberthis
  \ .
\end{align*}
For the stress-energy tensor, which has a background value of
\begin{equation}
 \bar{T}_{\mu\nu}=\bar{\rho}\ \bar{u}_\mu \,\bar{u}_\nu+\bar{p}_r\, \bar{r}_\mu \,\bar{r}_\nu+ \bar{p}_\varphi\left( \bar{g}_{\mu\nu}+\bar{u}_\mu \,\bar{u}_\nu-\bar{r}_\mu \,\bar{r}_\nu\right)
    \ ,
\end{equation}
where
\begin{equation}
  \begin{split}
    &\bar{u}_\mu= \left(-\sqrt{f(r)},\ 0,\ 0,\ 0\right), \quad \bar{r}_\mu= \left(0,\ 1/\sqrt{f(r)},\ 0,\ 0\right),\quad \\
  &\bar{\rho}=-\bar{p}_r=\frac{1-f(r)-r\, f'(r)}{r^2},\quad \bar{p}_\varphi=\frac{2f'(r)+r\,f''(r)}{2r}
  \ ,
  \end{split}
\end{equation}
the dependence on the metric is implicit, so one can calculate the perturbations on general grounds by imposing the variation
\begin{equation}
  \begin{split}
    &\hspace{1cm}u_\mu=\bar{u}_\mu+\delta u_\mu,\quad r_\mu=\bar{r}_\mu+\delta r_\mu,\\
    &\rho=\bar{\rho}+\delta\rho,\quad p_r=\bar{p}_r+\delta p_r,\quad p_\varphi=\bar{p}_\varphi+\delta p_\varphi
    \ ,  
  \end{split}
\end{equation}
where the perturbations $\delta u_\mu$ and $\delta r_\mu$ can be fixed via the perturbed orthonormality conditions:
\begin{align}
  \label{eq:perturbedorthonorm}
  \delta u_\mu\, \bar{u}^\mu=\frac{h^{\mu\nu}\bar{u}_\mu \bar{u}_\nu}{2},\quad \delta r_\mu\, \bar{r}^\mu=\frac{h^{\mu\nu}\bar{r}_\mu \bar{r}_\nu}{2},\quad \delta u_\mu\, \bar{r}^\mu+\delta r_\mu\, \bar{u}^\mu=h^{\mu\nu}\bar{u}_\mu \bar{r}_\nu
  \ ,
\end{align}
and $\delta \rho$, $\delta p_r$ and $\delta p_\varphi$ can be fixed via the perturbed conservation equation:
\begin{equation}
  \label{eq:perturbedconservation}
  \bar{\nabla}^\mu\left(\left.\frac{\delta T_{\mu\nu}}{\delta g_{\rho\sigma}}\right|_{\bar{g}} h_{\rho \sigma}\right)=\bar{\nabla}^\mu\left(\tensor{h}{_\mu^\rho}\,\bar{T}_{\rho\nu}\right)+\frac{1}{2}\bar{\nabla}_\nu\left(h^{\rho\sigma}\right)\bar{T}_{\rho\sigma}-\frac{1}{2}\bar{\nabla}^\mu\left(\tensor{h}{_\rho^\rho}\right)\bar{T}_{\mu\nu}
  \ .
\end{equation}
Given these conditions, we obtain a physically sensible perturbation of the stress-energy tensor, which takes then the following form:
\begin{equation}
  \begin{split}
    \left.\frac{\delta T_{\mu\nu}}{\delta g_{\rho\sigma}} \,\right|_{\bar{g}} h_{\rho \sigma}=&\ \delta\rho\ \bar{u}_\mu \,\bar{u}_\nu+\delta p_r\, \bar{r}_\mu \,\bar{r}_\nu+ \delta p_\varphi\left( \bar{g}_{\mu\nu}+\bar{u}_\mu \,\bar{u}_\nu-\bar{r}_\mu \,\bar{r}_\nu\right)\\
  &\ +2\,\bar{\rho}\,\delta{u}_{(\mu} \,\bar{u}_{\nu)}+2\, \bar{p}_r\,\delta{r}_{(\mu} \,\bar{r}_{\nu)} + \bar{p}_\varphi\left( h_{\mu\nu}+2\,\delta{u}_{(\mu} \,\bar{u}_{\nu)}-2\,\delta{r}_{(\mu} \,\bar{r}_{\nu)}\right)
  \ ,  
  \end{split}
\end{equation}
Now, for the study of the odd gravitational perturbations, which are algebraically simpler, we impose the  Regge-Wheeler gauge \cite{reggewheeler} and set $h_2=0$ in (\ref{eq:decompositiongravodd}). From (\ref{eq:perturbedorthonorm}) we get
\begin{equation}
  \delta u_{\mu}^{\text{(o)}}=\delta r_{\mu}^{\text{(o)}}=0
  \ ,
\end{equation}
while from Eq. (\ref{eq:perturbedconservation}) we get
\begin{align}
  &\delta \rho^{\text{(o)}}=\delta p_\varphi^{\text{(o)}}=0\\
  &\delta p_r^{\text{(o)}}=e^{-i\omega t}\, Y_{\ell m}\,\frac{\epsilon}{\sqrt{f(r)}r^2}
  \ ,
\end{align}
where $\epsilon$ is an arbitrary constant of integration that comes out of the conservation equation, which by consistency must be in the order of the perturbations.
After writing down the explicit form of the perturbed Einstein equations and relabeling 
\begin{equation}
  \psi_2^{\rm (o)}=\frac{f(r)\, h_1}{r}
  \ ,
\end{equation}
we get the Schr\"odinger-like equation (\ref{eq:schrodinger}) with the potential in Eq. (\ref{eq:oddgravpertpot}).

For the even gravitational perturbations the gauge choice is not as simple anymore. We will fix $F_0=0$, while $F_1$ and $G$ will be implicitly defined such as to satisfy the following equations:
\begin{equation}
  \label{eq:gaugeevenG}
  G=\frac{4f(r)}{\ell(\ell+1)rf'(r)}\biggl[\ell(\ell+1)\left(\frac{F_1}{r}-\frac{r}{2}\frac{dG}{dr}\right)+H_0-H_2+r\frac{dK}{dr}+\frac{rf'(r)}{2f(r)}K\biggr]
\end{equation}
\begin{equation}
  \label{eq:gaugeevenF1}
  \begin{split}
    &F_1=\frac{rK}{2f(r)}-\frac{\Tilde{\epsilon}\,r}{f(r)^{3/2}\,(2-2f(r)+r^2f''(r))}\\
  &\hspace{1cm}+\frac{rG}{4f(r)(\ell(\ell+1)-2f(r)+rf'(r))(2-2f(r)+r^2f''(r))}\\
  &\hspace{1.6cm}\times\Bigl\{\left[2+r^2f''(r)\right]\left[(\ell(\ell+1))^2-\ell(\ell+1)rf'(r)-2r^2f'(r)^2\right]\\
  &\hspace{2.4cm}+4f(r)^2\left[\ell(\ell+1)-2rf'(r)+r^3f'''(r)\right]\\
  &\hspace{2.4cm}-2f(r)\bigl[-4r^2f'(r)^2+(-4+3\ell(\ell+1))\,r^2f''(r)\\
  &\hspace{4cm}-2r^4f''(r)^2+\ell(\ell+1)(\ell(\ell+1)+2+r^3f'''(r))\\
  &\hspace{4cm}+rf'(r)(-3\ell(\ell+1)+2r^2f''(r)+r^3f'''(r))\bigr]\Bigr\}  
  \ ,
    \end{split}
\end{equation}
where $\Tilde{\epsilon}$ is an arbitrary constant for now. As we did for the odd perturbations, we can
now calculate $\delta u_\mu$, $\delta r_\mu$, $\delta\rho$, $\delta p_r$ and $\delta p_\phi$
with this specific form of the perturbation. A short calculation with Eq.~\eqref{eq:perturbedorthonorm}
reveals that
\begin{align}
   &\delta u_{\mu}^{\text{(e)}}= e^{-i\omega t}\, Y_{\ell m}\left(\frac{\sqrt{f(r)}\,H_0}{2},\ \frac{H_1}{2\,\sqrt{f(r)}},\ 0,\ 0\right)
\end{align}
\begin{align}
   &\delta r_{\mu}^{\text{(e)}}=e^{-i\omega t}\, Y_{\ell m}\left(\frac{\sqrt{f(r)}\,H_1}{2},\ \frac{H_2}{2\,\sqrt{f(r)}},\ 0,\ 0\right)\
   \ ,
\end{align}
while from~\eqref{eq:perturbedconservation} we get that
\begin{align}
  &\delta p_\varphi^{\text{(e)}}=e^{-i\omega t}\, Y_{\ell m}\,\frac{2-2f(r)+r^2f''(r)}{2r^3}\left(H_0-H_2\right)\\
  &\delta\rho^{\text{(e)}}=e^{-i\omega t}\, Y_{\ell m}\,\frac{2-2f(r)+r^2f''(r)}{2r^2}\left[\frac{\ell(\ell+1)}{2}G-K\right]\\
  \label{eq:deltapreven}
  &\delta p_r^{\text{(e)}}=e^{-i\omega t}\, Y_{\ell m}\,\frac{\epsilon}{\sqrt{f(r)}r^2}
    \ ,
\end{align}
where again $\epsilon$ is a constant of integration in the order of the perturbations.
This result is highly non-trivial, because now, for the even perturbations, the conservation
equation~\eqref{eq:perturbedconservation}) prescribes a rather involved differential equation
for $\delta p_r$, where the matter degrees of freedom get mixed with the gravitational
perturbations ones.
The gauge choice for $G$ in Eq.~\eqref{eq:gaugeevenG} has the precise effect of decoupling
the two different degrees of freedom and producing an integrable differential equation for $\delta p_r$,
which results in the simple form in Eq.~\eqref{eq:deltapreven}.
Since for the previous calculations the value of $\Tilde{\epsilon}$ in Eq.~\eqref{eq:gaugeevenF1}
was irrelevant, we may fix it now to be $\Tilde{\epsilon}=\epsilon$. 
\par
The process of extracting now the Schr\"odinger-like equation from the perturbed Einstein
equations is straightforward, even though tedious. Motivated by \cite{gravmoncrief},
we introduce the relabelling
\begin{equation}
\psi_2^{\rm (e)}
=
r\,K+\frac{2f(r)}{\ell(\ell+1)-2f(r)+rf'(r)}
\left[rH_2-r^2\frac{dK}{dr}+\ell(\ell+1)\left(\frac{r^2}{2}\frac{dG}{dr}-F_1\right)\right]
\ ,
\end{equation} 
which is the variable that will appear in Eq.~\eqref{eq:schrodinger}.
One eventually arrives at the potential given in Eq.~\eqref{eq:evengravpertpot}, 
with the gauge condition~\eqref{eq:gaugeevenF1} playing a crucial role for simplifying
the final expression.
The potential in Eq.~\eqref{eq:evengravpertpot} is the generalization in our context of the Zerilli
potential for a Schwarzschild black hole~\cite{gravmoncrief, zerilli}.
\section*{}
%
%
%
\bibliography{references}
\bibliographystyle{utphys}
\end{document}